\documentclass[utf8]{frontiersSCNS} 

\pdfoutput=1
\usepackage{lineno,microtype,subcaption}
\usepackage[hyphens]{url}
\usepackage{hyperref}
\usepackage{breakurl}
\usepackage[onehalfspacing]{setspace}
\usepackage{graphicx}
\usepackage{multirow}
\usepackage{colortbl}
\usepackage{xcolor}
\usepackage{pgfplots}
\pgfplotsset{compat=1.9}
\usetikzlibrary{plotmarks}
\usepackage{wasysym}

\def\keyFont{\fontsize{8}{11}\helveticabold }
\def\firstAuthorLast{Myriam Alanis-Espinosa {et~al.}} 
\def\Authors{Myriam Alanis-Espinosa\,, David Guti{\'e}rrez\,$^{*}$}
\def\Address{Laboratory of Biomedical Signal Processing, Center for Research and Advanced Studies (Cinvestav) at Monterrey, Apodaca , Mexico}
\def\corrAuthor{David Guti{\'e}rrez}

\def\corrEmail{dgtz@ieee.org}

\begin{document}
\onecolumn
\firstpage{1}

\title[Brain connectivity associated to an immersive BCI]{On the assesment of functional connectivity in an immersive brain-computer interface during motor imagery}

\author[\firstAuthorLast ]{\Authors} 
\address{} 
\correspondance{} 

\extraAuth{}

\maketitle

\begin{abstract}


New trends on brain-computer interface (BCI) design are aiming to combine this technology with immersive virtual reality in order to provide a sense of realism to its users. In this study, we propose an experimental BCI to control an immersive telepresence system using motor imagery (MI). The system is immersive in the sense that the users can control the movement of a NAO humanoid robot in a first person perspective (1PP), i.e., as if the movement of the robot was his/her own. We analyze functional brain connectivity between 1PP and 3PP during the control of our BCI using graph theory properties such as degree, betweenness centrality, and efficiency. Changes in these metrics are obtained for the case of the 1PP, as well as for the traditional third person perspective (3PP) in which the user can see the movement of the robot as feedback. As proof-of-concept, electroencephalography (EEG) signals were recorded from two subjects while they performed MI to control the movement of the robot. The graph theoretical analysis was applied to the binary directed networks obtained through the partial directed coherence (PDC). In our preliminary assessment we found that the efficiency in the $\alpha$ brain rhythm is greater in 1PP condition in comparison to the 3PP at the prefrontal cortex. Also, a stronger influence of signals measured at EEG channel C3 (primary motor cortex) to other regions was found in 1PP condition. Furthermore, our preliminary results seem to indicate that $\alpha$ and $\beta$ brain rhythms have a high indegree at prefrontal cortex in 1PP condition, and this could be possibly related to the experience of sense of agency. Therefore, using the PDC combined with graph theory while controlling a telepresence robot in an immersive system may contribute to understand the organization and behavior of brain networks in these environments.

\tiny
 \keyFont{ \section{Keywords:} Brain-computer interface, Partial directed coherence, graph theory, sense of agency, functional brain connectivity} 

\end{abstract}

\section{Introduction}


A brain-computer interface (BCI) is a system that enables a real-time user-device communication
pathway through different types of brain activity. 
In the beginning, BCIs were aimed for people with a disability of motor control or speech. Nowadays, even healthy people is using this technology for different applications. Some examples of BCI with immersive applications include gaming~\citep{Lalor.2005}, training~\citep{Vourvopoulos.2016}, rehabilitation~\citep{Calabro.2017}, or psychological treatments~\citep{Jancke.2009}. Furthermore, there have been some applications in which a user can teleoperate a robot, i.e., the user can have control of a robot that is not placed in the same location as him/her. Some examples of such applications are shown in~\cite{Escolano.2012}, \cite{Leeb.2015}, \cite{Beraldo.2018}. In these cases, the subject perceives the environment real and in 3D as a extension of his/her sensorial functions. Such extension increases the feeling of presence of a remote scenario as well as a sense of agency when moving~\citep{ref1}. This can be achieved by using either technologies based on head-mounted display (HMD) or multiple projections. Immersive virtual reality (VR) can also use HMD to project the virtual space just in front of the eyes, then the users focus on the display without distraction. 

In the last years, there has been an increase in the number of research about sense of presence, embodiment, and sense of agency when combining BCI and immersive environments. Nevertheless, most studies of sense of presence have focused in the subjective experience analysis through questionnaires, like in~\cite{Friedman.2007}, \cite{Baka.2015b}. Only a few studies involve the analysis of brain activity with the aim to explain cognitive processes related to the sense of agency in VR environments~\citep{Baumgartner.2008}.
Therefore, the purpose of this study is to further contribute to this area with a quantitative analysis of the brain connectivity while controlling a BCI teleoperated robot in an immersive environment, in our case through graph theory metrics. For that purpose, we first describe the BCI which is controlled with motor imagery (MI) and two conditions: the first person perspective (1PP), i.e., the immersion experience, and the traditional third person perspective (3PP) of visual feedback. 
Next, we introduce the partial directed coherence (PDC) as the metric that allows us to assess the functional brain connectivity by calculating a connectivity matrix based on electroencephalography (EEG) measurements at various brain frequency bands: $\theta$ (4-7~Hz), $\alpha$ (8-13~Hz), $\beta$ (14-29~Hz), and $\gamma$ (30-50~Hz). 

We already showed the usefulness of the PDC for the analysis of functional brain connectivity in~\cite{gaxiola}, \cite{alanis}. Yet, there is still work to do in how to interpret the interactions that the PDC reveals. The brain functional connectivity represents a complex system because of the transient nature of the interactions, such as synchronizations and desynchronizations between different brain regions. Therefore, it is complicated to compare the cortical connections within different brain rhythms. For this reason, here we propose the use of graph theory to describe the involvement of the different EEG channels (based on the PDC values) over the frontal, central, parietal, and occipital regions. Graph theory is a powerful tool for understanding the interactions and topology of various types of networks, and it 
has found a place in neurosciences for investigating brain aging~\citep{Vecchio.2014}, and different brain disorders like in mild cognitive impairment~\citep{Berlot.2016}, and brain function in spinal cord injured patients~\citep{VicoFallani.2007}. In BCI research, graph theory has been already proposed to analyze brain networks of different mental tasks~\citep{Huang.2016}, \citep{StefanoFilho.2018}. 
In this paper, we propose the use of a series of graph theory metrics in order to understand the differences in functional brain connectivity that arise depending on the immersion experience that our users have with our proposed BCI system.

\section{Materials and Methods}

As proof-of-concept, we acquired the EEG recordings from two healthy right-handed participants (25 years-old female and 24 years-old male) that were asked to operate our proposed BCI system. Participants were recruited from the \emph{Center of Research and Advanced Studies at Monterrey} and they were not familiar with BCI's. Informed consent was obtained from the participants after explanation of the study.

The experimental protocol consists of two stages. The first one is the conventional BCI training (2D, monitor-based), in which imaginary movements are performed according to the visual cues presented without feedback. The second stage corresponds to the BCI control task. Next, we will explain both of them.

\subsection{Conventional BCI training}\label{sec:training}
\begin{figure}[h]
	\centering
		\includegraphics[width=0.50\textwidth]{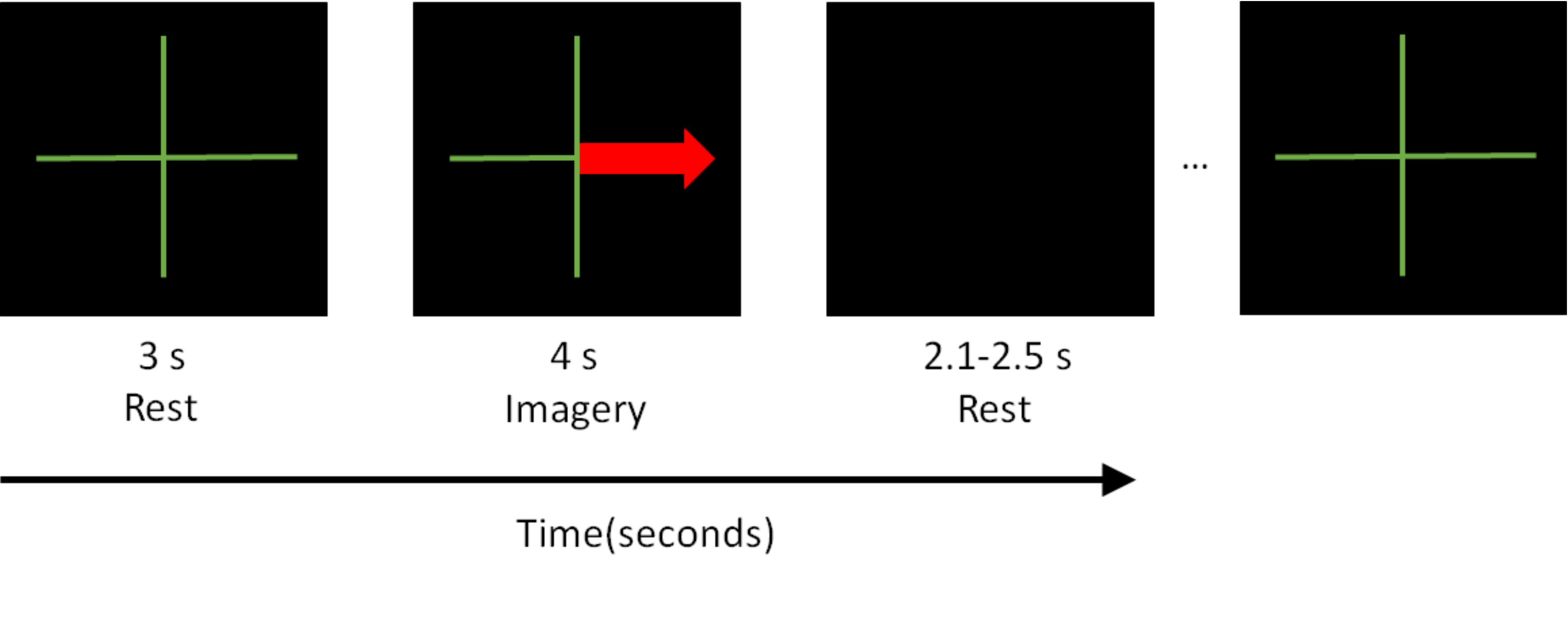}
	\caption{Training sequence as shown in the computer monitor to the subject.}
	\label{fig:Training_sequence}
\end{figure}

The aim of this stage is to extract characteristic features from the EEG recordings to be used for the control of the BCI. First, the 
subject is required to perform imaginary movements or to remain in resting state in response to the visual cues presented on a computer monitor while sitting on a chair. The cues are in the form of red arrows pointing either left, right, up, or down. They indicating whether the subject has to imagine closing and opening either the left hand, right hand, both hands, or moving both feet, respectively. Hence, the subjects are tested for four different control conditions or \textit{classes}. Each trial starts with a fixation cross for 3~s in which the subject has to try to be with at rest, followed by the visual cue presented during 4~s. Each trial has a random duration interval of 2.1 to 2.5~s between them to avoid adaptation~\citep{Friedman.2007}. During this interval, the subject remains at rest. The training sequence is shown in Figure~\ref{fig:Training_sequence}. 

The subject has five minutes of rest between each one of the five runs. Each run corresponds to the acquisition of EEG measurements of forty trials (ten randomized presentations of the cues of each class). These trials are used as \emph{calibration} recordings to build a classifier for discriminating two different mental tasks (in our case, each of the tested classes compared against the resting state). The task that shows highest separability versus resting state is then used as the user-specific input for the BCI control task system in the next stage of the experiments.

\subsection{BCI control task}

\begin{figure}[h]
	\centering
		\includegraphics[width=0.50\textwidth]{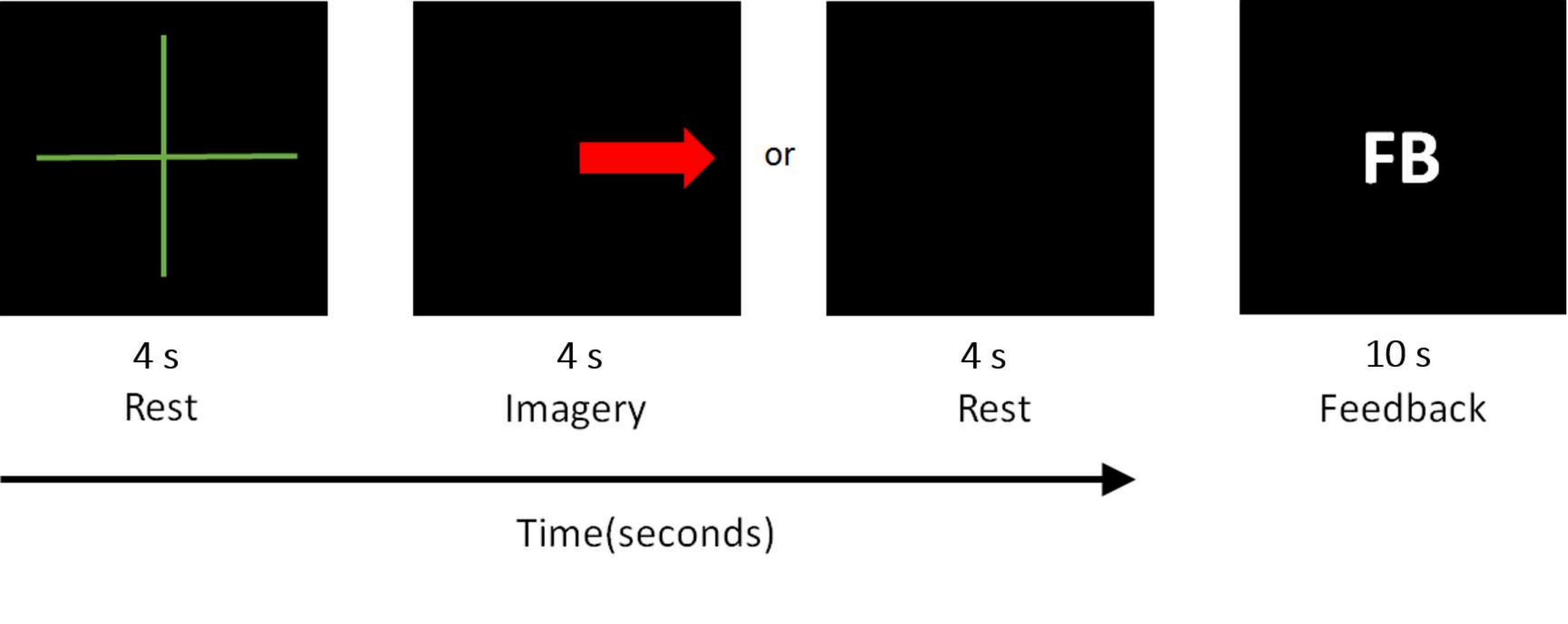}
	\caption{Control sequence as shown in the computer monitor and HMD.}
	\label{fig:Control_sequence}
\end{figure}

\begin{figure}[hb]
	\centering
		\includegraphics[width=0.35\textwidth]{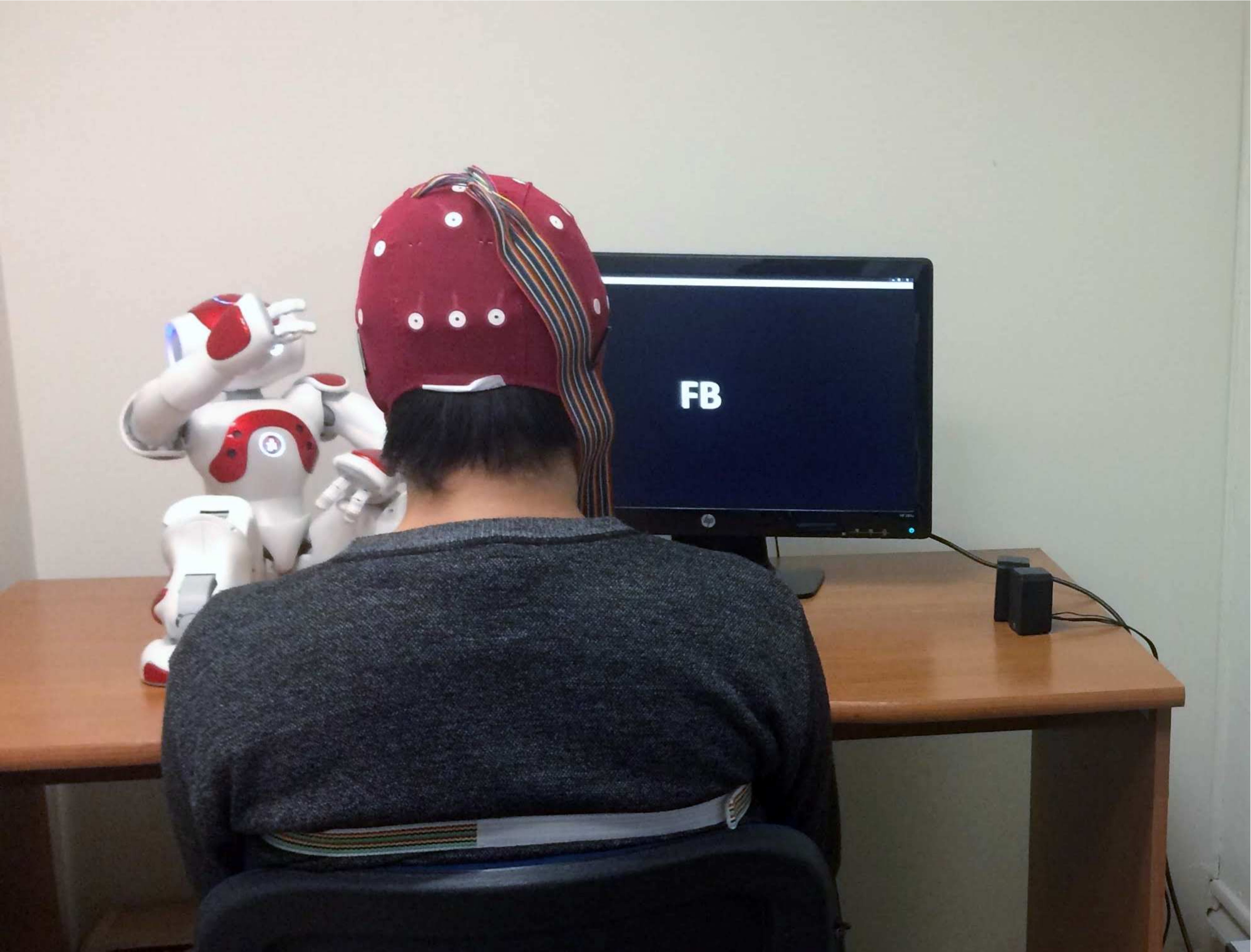}
		\caption{Third-person perspective.}
		\label{fig:3PP}
\end{figure}

For this stage, we used the NAO humanoid robot as the device the BCI system is to control. The main goal of the subject is to control the movement of the robot with the personalized classifier obtained in the training stage. The sequence displayed on the screen to subjects during the experiment is shown in Figure~\ref{fig:Control_sequence}. At the beginning of the experiment, the subject remains in resting state for 15~s. Then, the following sequence is displayed on the screen: it starts with a green cross, followed by a black screen or a red arrow, and finally the letters FB (for \emph{feedback}) appear. The first and second screen appear for 4~s, and the FB screen appears for 10~s with a random time between each trial from 100~ms to 500~ms with a black screen. During the FB screen, the subject receives feedback from the robot moving or not according to the cue. When a MI task is detected, the robot closes and opens its hand, otherwise, the robot does not move when resting state is detected. The two types of feedback implemented are the following: 

\textsl{3PP}: The subject is comfortably sitting in front of the robot and a monitor where the sequence of stimuli is shown, as it can be seen in Figure~\ref{fig:3PP}. In this case, the subject sees the robot moving in third-person perspective. For all trials, the robot is placed on the left side, and the monitor on the right side. 

\begin{figure}[hb!]
\begin{minipage}[b]{0.45\linewidth}
\centering
\includegraphics[width=0.8\textwidth]{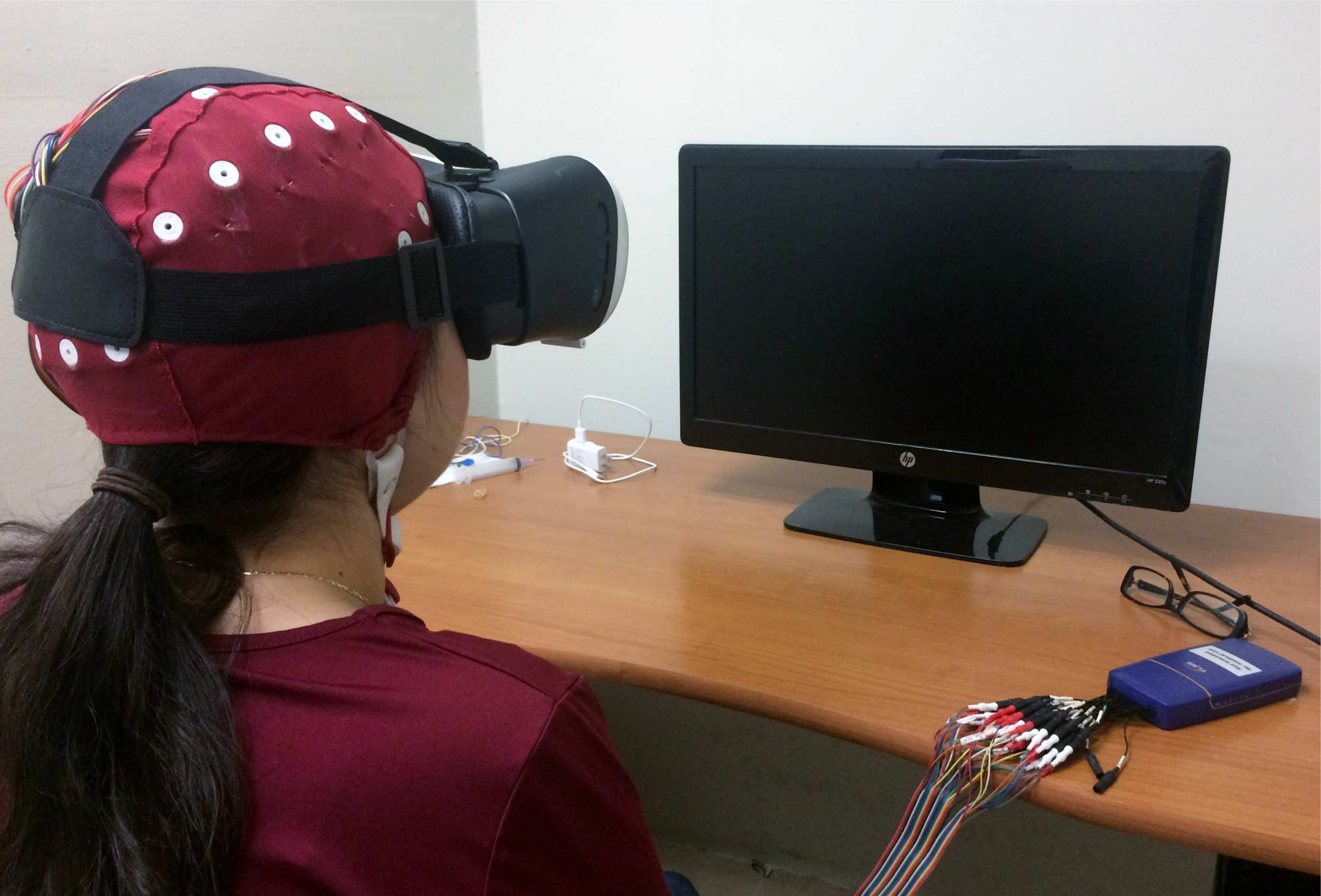}
\caption{First-person perspective.}
\label{fig:1PP}
\end{minipage}
\hspace{0.5cm}
\begin{minipage}[b]{0.45\linewidth}
\centering
\includegraphics[width=0.4\textwidth]{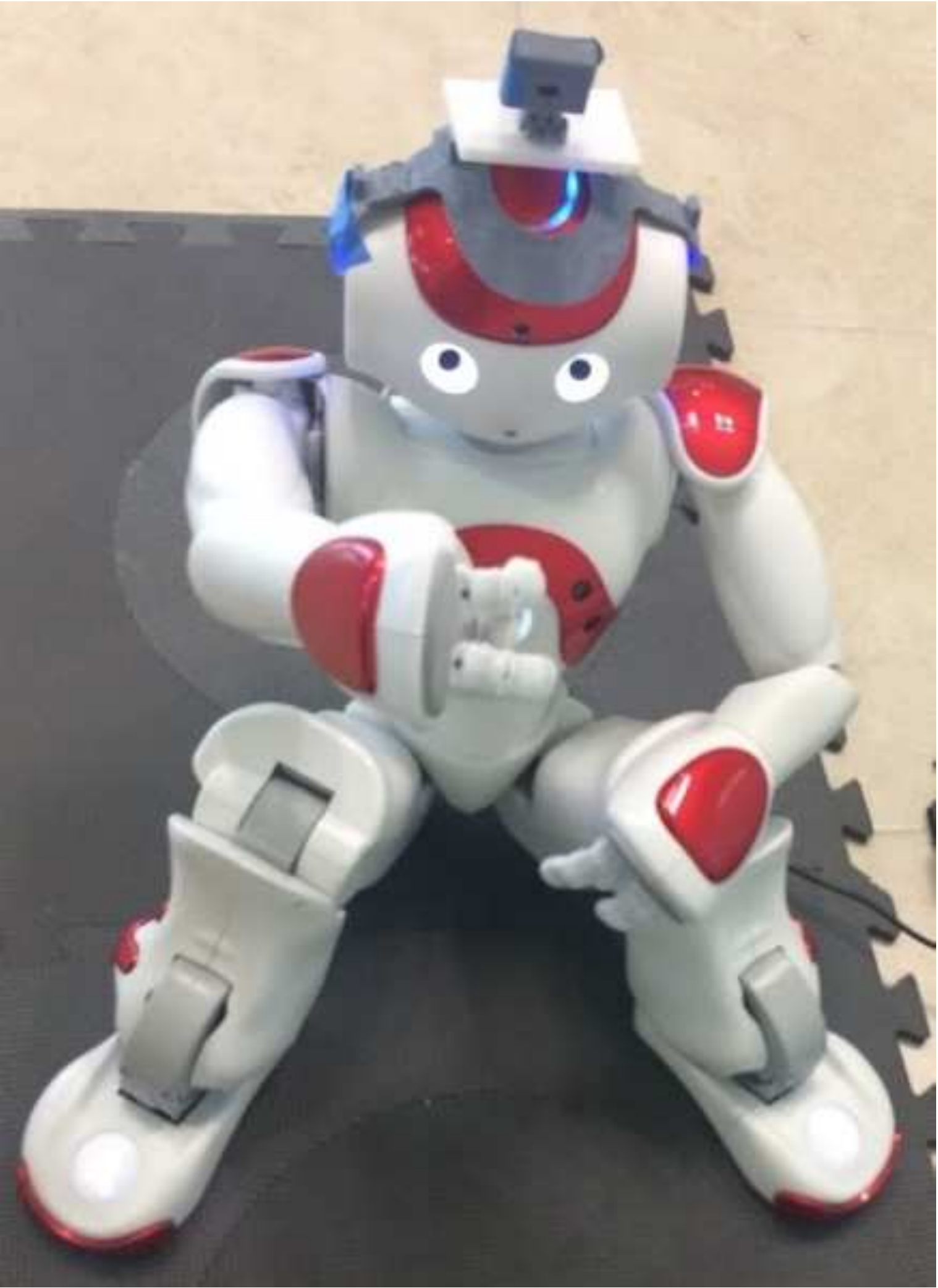}
\caption{NAO robot in another room.}
\label{fig:Nao}
\end{minipage}
\end{figure}

\textsl{1PP}: The subject is comfortably sitting using a passive HMD in which it is shown whatever the robot sees, as well as the stimuli sequence (see Figure~\ref{fig:1PP}). Meanwhile, the robot is placed outside the room as shown in Figure~\ref{fig:Nao}. The implementation of this feedback scheme is further detailed in Section~\ref{sec:ImmersiveSystem}. 

The subject is instructed to avoid blinking during the MI task to minimize noise in EEG data. Additionally, the subject has five minutes of rest between runs. A total of 200 trials of 18-s of duration were recorded in 2 different sessions per perspective. The accuracy is recorded for each cue, which corresponds to the number of times that the subject correctly controlled the movement of the robot, as well as the number of times the robot halted during the subject's resting state.


\subsection{Immersive telepresence system}
\label{sec:ImmersiveSystem}

The aim of this system is to allow the BCI user to see the world through the perspective of the NAO robot, i.e., as if the robot's movements were the user's own. In order to be immersive, some requirements have to be considered in the design of the telepresence:
\begin{enumerate}
	\item BCI user must control the NAO robot placed in a remote site using the MI paradigm;
	\item control signals have to be sent wirelessly;
	\item control signals and video feedback must have the minimum lag for an immersive experience;
	\item BCI user has a stereoscopic video image as feedback displayed into a HMD.
\end{enumerate}

\begin{figure}[h!]
	\centering
		\includegraphics[width=\textwidth]{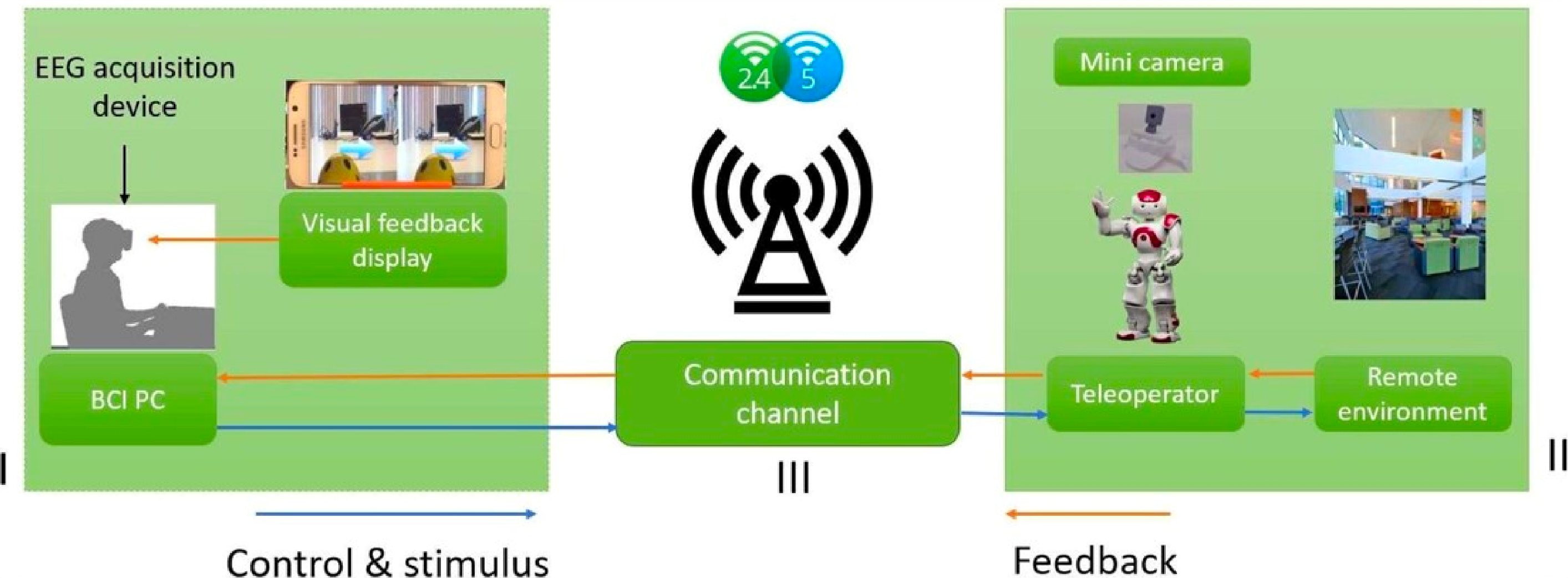}
	\caption{General framework for the implementation of the immersive telepresence system.}
	\label{fig:General_Framework}
\end{figure}

The implementation of the immersive telepresence system can be divided into two parts. The first contains the components of the BCI system considering the stereoscopic video feedback from the remote environment. The second part comprises the BCI software implementation, the system for the communication channel, and the system implemented at the remote environment. The general setup can be seen in Figure~\ref{fig:General_Framework}.

The hardware of the immersive telepresence system can be divided in three different modules:
\begin{itemize}
\item Module I corresponds to the location of both the BCI system and its user. It consists of a desktop computer (Intel core i3, 8 GB RAM) and the signal acquisition system MOBITA, which is a wireless EEG system with 32-channel. These two pieces are connected via 2.4~GHz wireless network. The virtual reality headset consisted of a passive HMD, coupled with a Samsung Galaxy S6 cellphone. The Galaxy S6 has a  large screen (1440 x 2560), an Exynos processor 7420 2.1 GHz, 3 GB of RAM and GPU Mali-T760MP8, and it is powerful enough to have an unnoticeable delay of the video streaming and VR web application which allows the immersion feeling.
\item In module II, at the NAO's robot site, a helmet with an Arducam of 5~MP and 1080p video resolution is placed on top of the robot's head in order to provide the required video feedback that the native robot's webcam cannot. The Arducam is connected directly to Raspberry Pi's native CSI camera port to provide better performance than a webcam in terms of the frame rate and resolution. 
\item Module III corresponds to the communication channel, which is implemented through an ASUS AC1200 router of double band and links to the main server running in the RPi 3.
\end{itemize}

\begin{figure}[h!]
	\centering
		\includegraphics[width=0.75\textwidth]{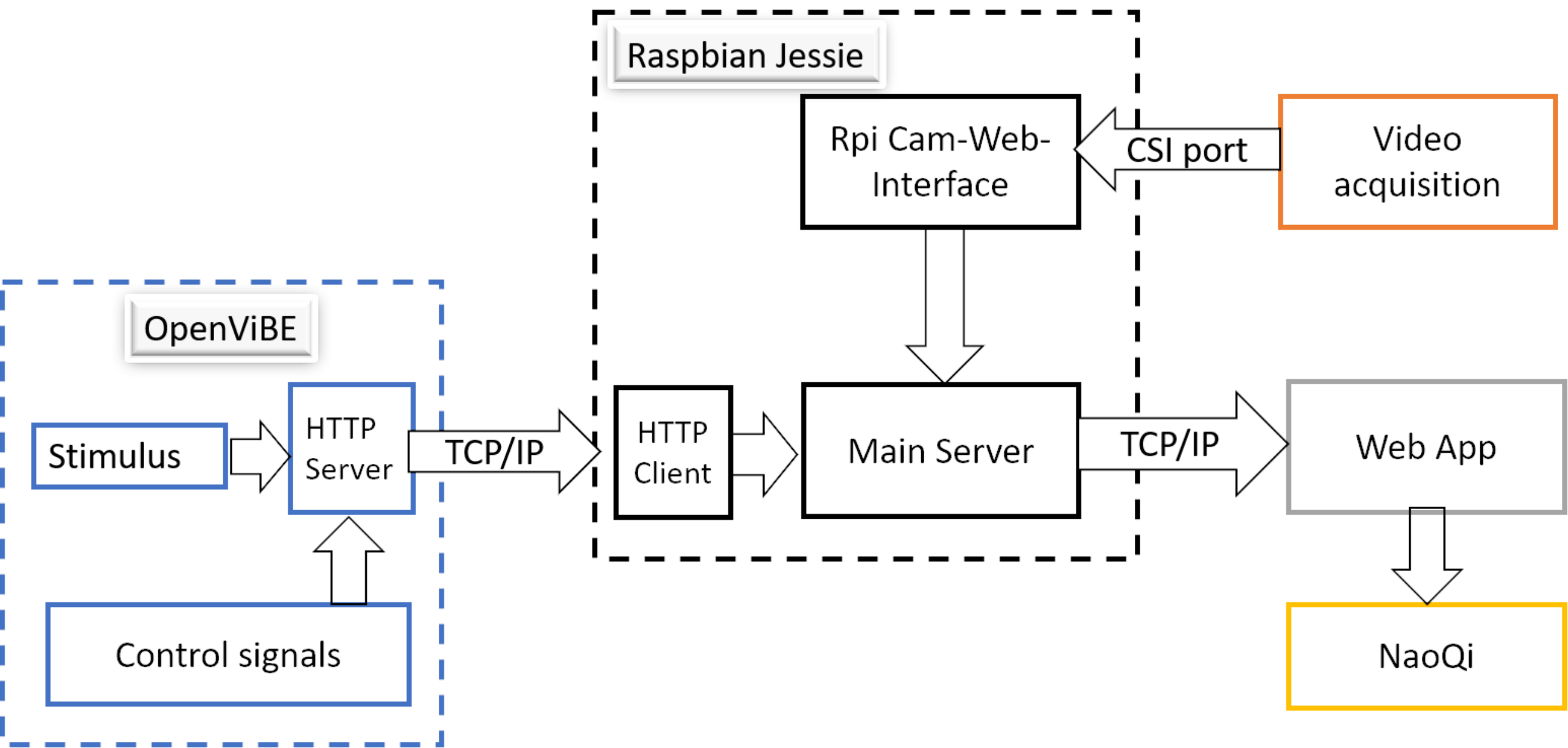}
	\caption{Software architecture}
	\label{fig:Software_architecture}
\end{figure}

The software architecture is shown in Figure~\ref{fig:Software_architecture}. In our case, the software used at the BCI user site is running in a desktop computer with Microsoft Windows 7 of 64 bits as operating system which runs Python 2.7 and OpenViBE~\cite{Renard.2010}. Within OpenViBE, the Python box is used to send the control signals to the RPi 3 and they are echoed to control robot using TCP/IP. 
Meanwhile, the stimulus provided by a LUA stimulator box is sent using an HTTP server written in Python (see \url{www.lua.org} for more details about LUA scripting language). The stimulus is received with a HTTP client at the RPi 3 and echoed to the main server.

At the teleoperator site, the RPi Cam-Web-Interface was implemented to manage the streaming video from the mini camera. Such interface is freely available at \url{https://github.com/silvanmelchior/RPi_Cam_Web_Interface}. In order to obtain a stereoscopic video image, 
we decided to use only the video from one camera as in the Pi Viewer demo available at \url{https://github.com/patcat/PiViewer}, in which the video from the RPi camera is streamed into VR using JavaScript. In our case, it was implemented with some modifications to enable the appearance of the stimulus images from OpenViBE without a noticeable delay into a web app.

\begin{figure}[h!]
	\centering
		\includegraphics[width=0.3\textwidth]{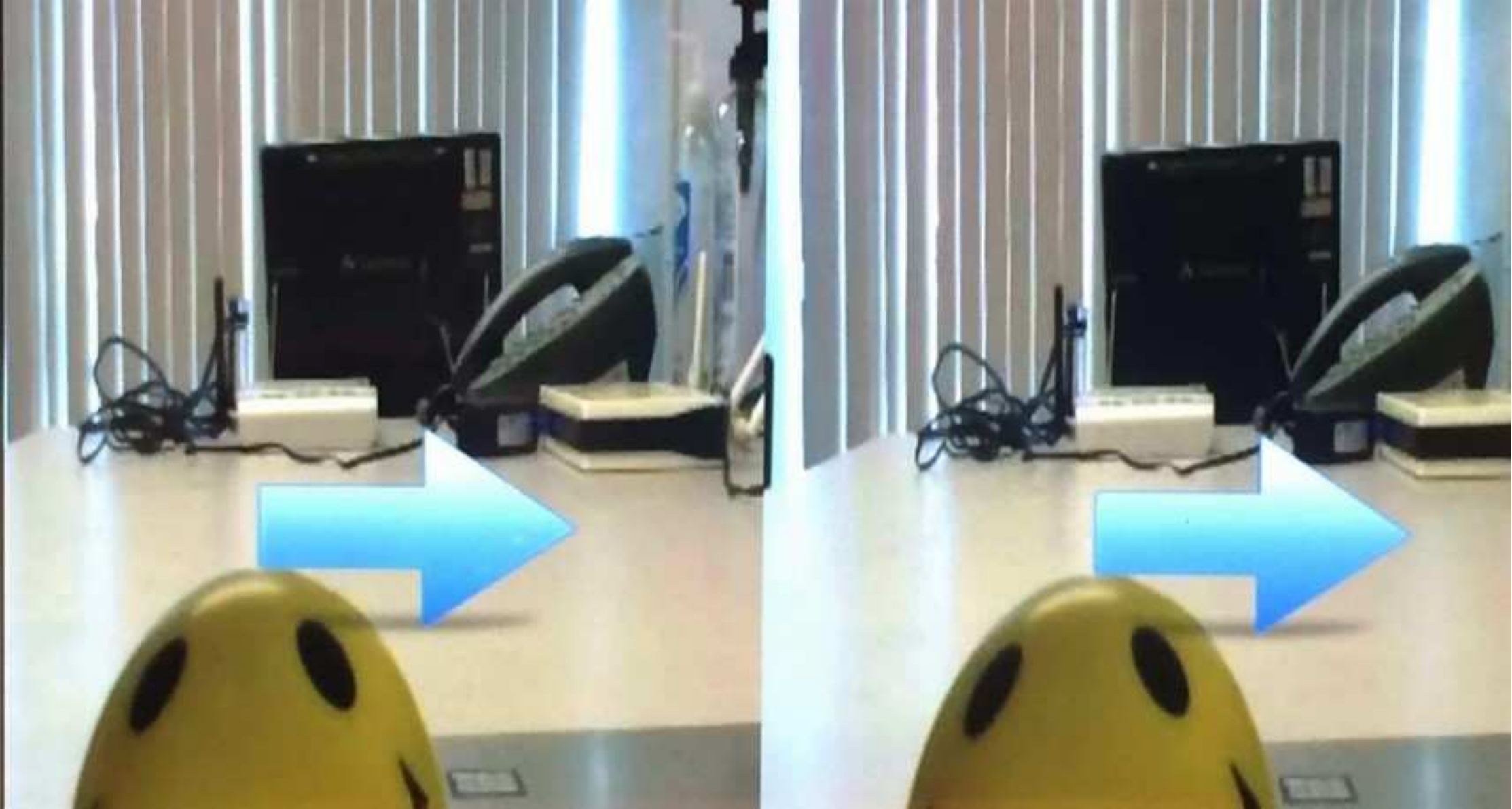}
	\caption{Stereoscopic image and BCI stimulus. The blue arrow is the stimulus to perform the motor imagery movement}
	\label{fig:VR_stimulus}
\end{figure}

The main server running in the RPi 3 was written with Node.js (\url{https://nodejs.org}). This server allows the integration of the camera and stimulus into a web app. It hosts the web page where it is shown a stereoscopic image of the remote environment streamed by the mini camera, as well as the different stimuli synchronized from the BCI user site, as it is shown in Figure~\ref{fig:VR_stimulus}. 

\subsection{EEG data acquisition and preprocessing}

Subjects were comfortably seated on a chair inside a noise free and normally lighted room.  A 32-channel EEG system (Mobita TMSi) was used to record the brain electrical potentials by means of an electrode cap with sensors placed according to the 10-20 international system and with reference to AFz. Impedance of all electrodes were kept below 5~k$\Omega$. The acquisition was performed at a sampling rate of 1000~Hz, the signals were bandpass filtered with a zero-phase fourth-order Butterworth between 1 to 100~Hz band and a notch filter to remove artifact caused by electrical power lines in 60~Hz. The blinking artifacts were removed using independent component analysis (ICA) and finally a baseline correction was performed.

\subsection{Feature Extraction and Classification}
The EEG signals obtained from the subjects after the conventional BCI training scheme described in Section~\ref{sec:training} were analyzed with the general-purpose software system BCI2000. This software allows for the calculation of personalized features for BCI control, and this is done by the calculation of the $r^2$ values for all electrodes within 0-70~Hz for two cognitive states. In this way, the channels and frequencies with higher $r^2$ values are selected as features to train the classifier. More details on this process can be found at ~\href{http://www.bci2000.org/mediawiki/index.php/User_Tutorial:Mu_Rhythm_BCI_Tutorial}{Mu Rhythm BCI Tutorial}. %

Once the features have been extracted, a linear discriminant analysis (LDA) classifier~\citep{LDA} is used to discriminate between the control command (detection of features that characterize the MI task) or the rest state.  

\subsection{Partial Directed Coherence}

The partial directed coherence (PDC) is a method based on the Granger causality to measure the coupling or connectivity between different channels in the frequency domain. The PDC also identifies the direction of information flow and the strength~\citep{Baccala.2001}. In our case, the frequency domains selected correspond to the known EEG rhythms $\theta$, $\alpha$, $\beta$ , and $\gamma$. The PDC was calculated only between the following EEG channels: FP1, FP2, F3, F4, Fz, C4, C3, Cz, C4, C3, Pz, T5, T6, O1, Oz and O2. We chose those channels based on previous work related to brain processing in the human visuo-motor system~\citep{Binkofski.2013}, and during motor imagery events~\citep{Ghosh.2015}. 

In order to compute the PDC, the data needs to be fitted to a multivariate autoregressive (MVAR) model. For the case of a set $S=\{x_m, 1\leq m\leq M \}$ of $M$ EEG signals (in our case those from the previously selected $M=16$ channels), the MVAR model of order $\rho$ of $\mathbf{x}(n)=[x_1(n), x_2(n), \ldots, x_i(n), \ldots, x_j(n), \ldots, x_M(n)]^T$, for $n=1, 2,\ldots, T$ time samples, is given by
\begin{equation}
	\mathbf{x}(n) = \sum_{p=1}^{\rho}A_p\mathbf{x}(n-p)+\mathbf{v}(n),
\end{equation}
where $A_1, A_2, \ldots, A_\rho$ are the $M\times M$ coefficient matrices containing the coefficients $a_{ij}(p)$ accounting for the linear interaction effect of $x_j(n-p)$ onto $x_i(n)$, and $\mathbf{v}(n)=[v_1(n), v_2(n), \ldots, v_M(n)]^T$ is the noise vector (uncorrelated error process). Under those conditions, a measure of the direct causal relations (directional connectivity) of $x_j$ to $x_i$ for a frequency $f$ is given by the PDC as

\begin{equation}
	\pi_{ij}(f) = 
	\frac{A_{ij}(f)}{\sqrt{\mathbf{a}_j(f)\mathbf{a}_j^T(f)}},
\label{eq:PDC}
\end{equation}
where $A_{ij}(f)$ and $\mathbf{a}_j(f)$ are, respectively, the $i, j$ element and the $j$-th column of
\begin{equation}
	\mathbf{A}(f) = \mathbf{I} - \sum_{p=1}^{\rho}A_p e^{-j2\pi fp}.
\end{equation}
In this work, the connectivity analysis is done during MI epoch from the first to the third second after the cue is presented in order to calculate the PDC values for each frequency between 4 and 50 Hz. All values are arranged in matrices of size $16\times 16$ (to account for all $\pi_{ij}$ values of our EEG channels) for each of the frequencies. Then, the maximum PDC value is calculated  between those PDC matrices corresponding to the frequencies of each brain rhythms. At the end, we are left with four PDC matrices per trial (representing $\theta$, $\alpha$, $\beta$ , and $\gamma$), which are then used for the graph analysis.

\subsection{Graph analysis}

Graph theory has been used to describe large scale networks in different research fields. In neuroscience, graph theory is employed as a network analysis to identify the simultaneous activity of different brain regions stimulated by a mental state. In our case, %
a \emph{brain network} is obtained based on the PDC matrix generated with the PDC values of each pair of channels $\{i,j\}$. From the perspective of graph theory, the set of \emph{nodes} correspond to the EEG channels, while the set of \emph{edges} represent an anatomical link between those nodes or a functional dependence~\citep{Sporns.2010}. These pairwise connections are accounted for in a connection matrix. 

In order to compute the topological features of our interest, the connectivity matrix is converted into a directed, unweighted graph (also referred to as \emph{digraph}). Such unweighted matrix is then binarized by choosing a threshold $\tau$ that represents the number of most powerful connections (connection density). In our study, $\tau$ represents the ratio between the effective connections and all the possible ones in the digraph. A range of thresholds per frequency band were explored: between 0.1 and 0.9 with increments of 0.05. Then, we identified the minimum value in which there was a statistically significant difference of the global efficiency metric when comparing it between the 1PP and 3PP cases. Once it was binarized, the digraph was characterized according to the degree, distribution degree, node betweenness centrality, and local efficiency. All those network metrics were computed using the brain connectivity toolbox for Matlab~\citep{Rubinov.2010}. Next, a more detailed description of each of them is presented.

\subsubsection{Degree distribution}
The degree $k$ of a node $i$ measures the number of connections, so it indicates how many nodes are connected to node $i$. The degree distribution $p(k)$ represents the probability to find a node $i$ with certain degree $k$. In our case, the degree distribution can be divided in two: the indegree (ID) denoted as $k_{\mathrm{in}}$, and the outdegree (OD) denoted as $k_{\mathrm{out}}$. Those represent the total number of connections incoming to a node and outgoing from a node, respectively. A large value for  $k_{\mathrm{in}}$ means that the node is influenced 
by a large number of different channels. A large value of $k_{\mathrm{out}}$ means a node has a large number of potential targets.

\subsubsection{Betweenness centrality}
In a graph structure, we can identify important nodes that often interact with many others as a way to facilitate functional integration~\citep{Rubinov.2010}. 
Within a graph, we can also find central nodes that participate in many short paths, and these act as important controllers of information flow. Betweenness centrality is a sensitive measure defined as the fraction of all shortest paths in the network that pass through a given node. This measure is used to identify important functional connections.

\subsubsection{Efficiency}
This metric was introduced by~\cite{Latora.2001} to measure how efficiently the nodes communicate between them. The efficiency $e_{ij}$ of the communication between the nodes $i$ and $j$ is inversely proportional to the shortest path length $d_{ij}$. If a path does not exist between the nodes $i$ and $j$, then $d_{ij}= \infty $ and $e_{ij}= 0$. The global efficiency is defined as 
\begin{equation}
	E_{\mathrm{global}}=\frac{1}{N(N-1)}\sum_{i\neq j}{\frac{1}{d_{ij}}},
\end{equation}
where $N$ corresponds to the number of nodes composing the graph.

\subsection{Statistic assessment of Connectivity} 

Based on eConnectome toolbox~\citep{He.2011}, a nonparametric method based on surrogate data is used to assess the statistical significance of our calculated values of PDC. The Fourier transform is applied to each trial of the original data to randomly shuffle the phases without changing the magnitude as follows.  Next, inverse Fourier transform is applied to each trial with permutated phase to generate surrogate data in time, followed by the PDC estimation. The shuffling and PDC calculation are repeated 1000 times, resulting in a distribution of PDC values under the null hypothesis that no connectivity exists and with a significance level of $0.05$. 

\section{Results}

The analysis within the two types of experiments was done with the MI of the right hand because it was the one that presented the highest values of $r^2$ for Participant 1 (P1) and Participant 2 (P2). Therefore, the movement of the NAO robot was chosen to be also the closing and opening of its right hand to promote the sense of agency. Once we have the connectivity matrix calculated with the PDC, the digraphs were obtained according to a specific threshold for each frequency band, as shown in Table~\ref{tab:thresholds}. 
\begin{table}[b!]
\centering
\begin{tabular}{@{} *5l @{}}    \toprule
\emph{Brain rhythms} &$\theta$&$\alpha$&$\beta$&$\gamma$\\ \\\midrule

Threshold $\tau$ & 0.65 & 0.3 & 0.65 & 0.9\\ 
$p$ value & 0.007 & 0.05 & 0.05 & 0.05\\ 
 \hline
\end{tabular}
	\caption{Thresholds used to calculate the digraphs for each frequency band.}
	\label{tab:thresholds}
\end{table}

\begin{figure}[t!]
	\centering
		\resizebox {0.95\textwidth} {!} {
	\input{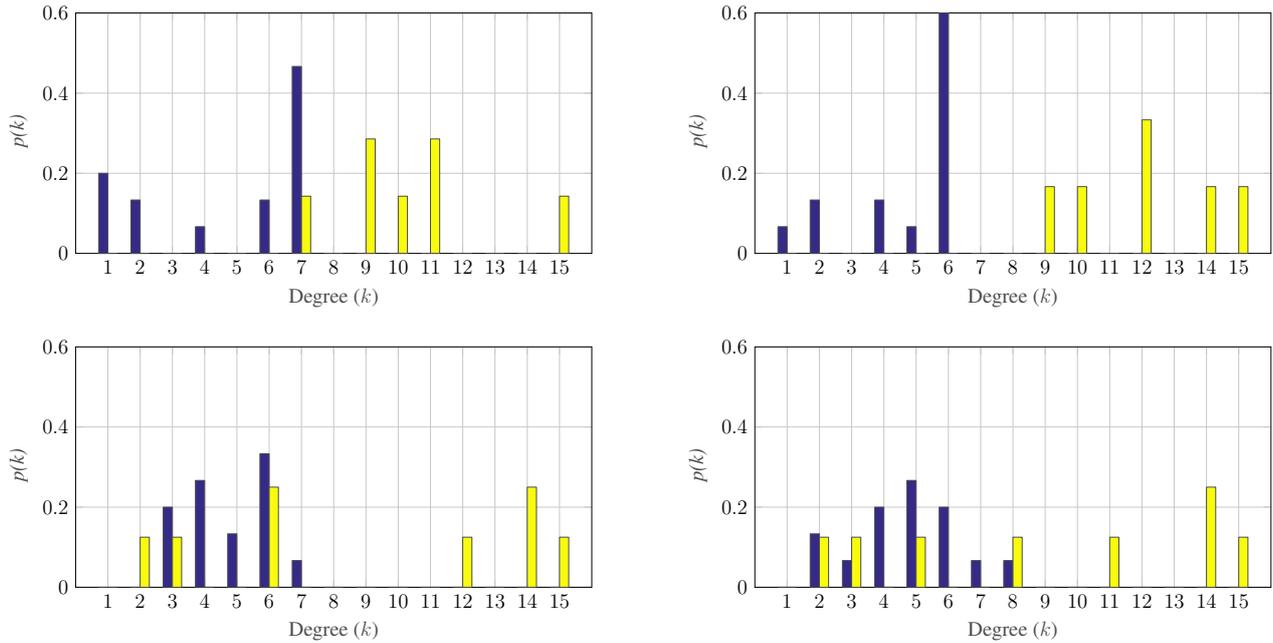}
	}
	\caption{Degree distributions of P1 (left) and P2 (right) shown for the 3PP (top) and 1PP (bottom) conditions in $\alpha$ frequency band. Indegree distribution is represented by dark-color bars and the outdegree is represented by light-color bars.}
	\label{fig:Distribution_alpha}
\end{figure}

Next, the degree distributions for ID and OD were calculated for each frequency band. Figure~\ref{fig:Distribution_alpha} shows the degree distribution of $\alpha$ for both participants and both the 1PP and 3PP conditions. Histograms are normalized to the number of possible nodes in the network (15 nodes). We note the consistency in the behavior of the distribution in both participants. An interesting result, is that the OD distribution shows different trends within each condition. The right-skewed tails of the OD distributions indicate that there are no nodes with less than 7 outgoing links. Contrary to the ID distribution, in which there are no nodes with more than 7 incoming connections. 

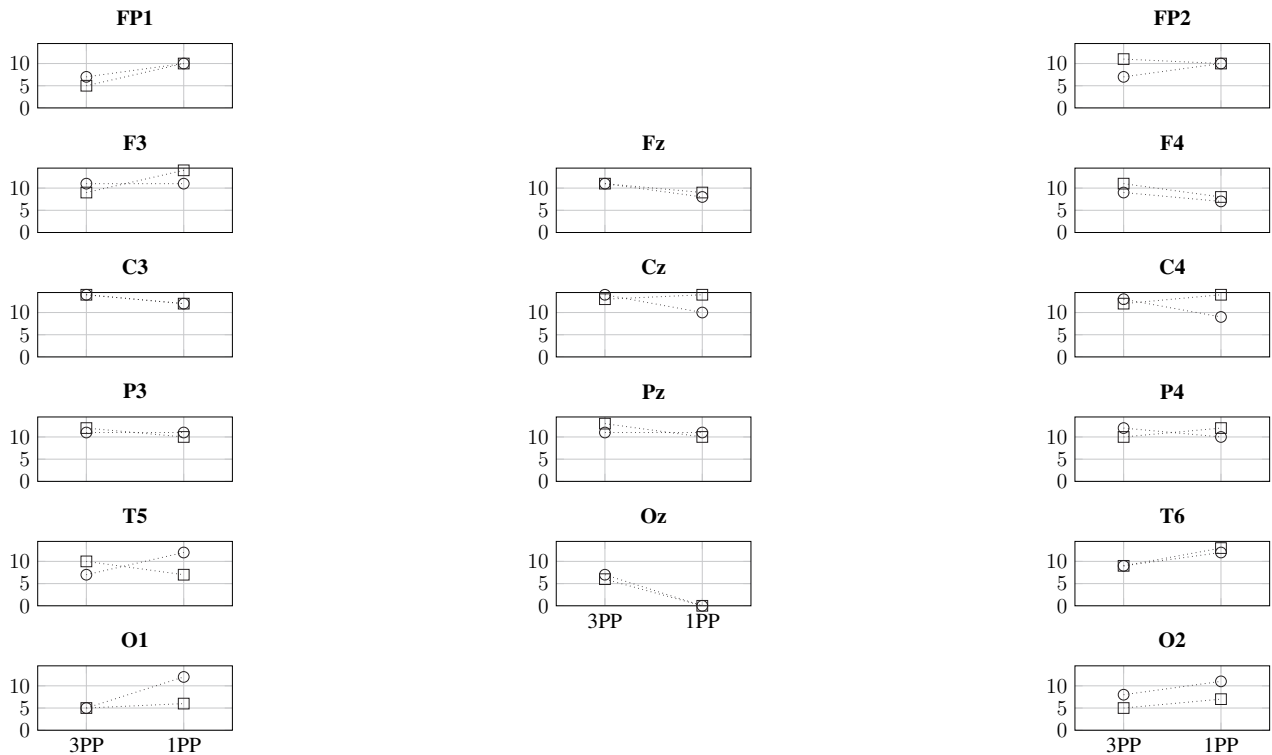
\begin{figure}
	\centering
	\resizebox {0.95\textwidth} {!} {
%
%
\begin{tikzpicture}

\begin{axis}[%
width=1.5in,
height=0.5in,
at={(1.85in,5.636in)},
scale only axis,
xmin=0.5,
xmax=2.5,
xtick={1,2},
xticklabels={{},{}},
ymin=0,
ymax=14.5,
axis background/.style={fill=white},
title style={font=\bfseries},
title={FP1},
xmajorgrids,
ymajorgrids
]
\addplot [color=black,  dotted, mark=o, mark size=3pt, mark options={solid, black}, forget plot]
  table[row sep=crcr]{%
1	7\\
2	10\\
};
\addplot [color=black,  dotted, mark=square, mark size=3pt, mark options={solid, black}, forget plot]
  table[row sep=crcr]{%
1	5\\
2	10\\
};
\end{axis}

\begin{axis}[%
width=1.5in,
height=0.5in,
at={(1.85in,4.669in)},
scale only axis,
xmin=0.5,
xmax=2.5,
xtick={1,2},
xticklabels={{},{}},
ymin=0,
ymax=14.5,
axis background/.style={fill=white},
title style={font=\bfseries},
title={F3},
xmajorgrids,
ymajorgrids
]
\addplot [color=black,  dotted, mark=o, mark size=3pt, mark options={solid, black}, forget plot]
  table[row sep=crcr]{%
1	11\\
2	11\\
};
\addplot [color=black, dotted, mark=square, mark size=3pt, mark options={solid, black}, forget plot]
  table[row sep=crcr]{%
1	9\\
2	14\\
};
\end{axis}

\begin{axis}[%
width=1.5in,
height=0.5in,
at={(1.85in,3.703in)},
scale only axis,
xmin=0.5,
xmax=2.5,
xtick={1,2},
xticklabels={{},{}},
ymin=0,
ymax=14.5,
axis background/.style={fill=white},
title style={font=\bfseries},
title={C3},
xmajorgrids,
ymajorgrids
]
\addplot [color=black,  dotted, mark=o, mark size=3pt, mark options={solid, black}, forget plot]
  table[row sep=crcr]{%
1	14\\
2	12\\
};
\addplot [color=black, dotted, mark=square, mark size=3pt, mark options={solid, black}, forget plot]
  table[row sep=crcr]{%
1	14\\
2	12\\
};
\end{axis}

\begin{axis}[%
width=1.5in,
height=0.5in,
at={(1.85in,2.737in)},
scale only axis,
xmin=0.5,
xmax=2.5,
xtick={1,2},
xticklabels={{},{}},
ymin=0,
ymax=14.5,
axis background/.style={fill=white},
title style={font=\bfseries},
title={P3},
xmajorgrids,
ymajorgrids
]
\addplot [color=black,  dotted, mark=o, mark size=3pt, mark options={solid, black}, forget plot]
  table[row sep=crcr]{%
1	11\\
2	11\\
};
\addplot [color=black, dotted, mark=square, mark size=3pt, mark options={solid, black}, forget plot]
  table[row sep=crcr]{%
1	12\\
2	10\\
};
\end{axis}

\begin{axis}[%
width=1.5in,
height=0.5in,
at={(1.85in,1.771in)},
scale only axis,
xmin=0.5,
xmax=2.5,
xtick={1,2},
xticklabels={{},{}},
ymin=0,
ymax=14.5,
axis background/.style={fill=white},
title style={font=\bfseries},
title={T5},
xmajorgrids,
ymajorgrids
]
\addplot [color=black,  dotted, mark=o, mark size=3pt, mark options={solid, black}, forget plot]
  table[row sep=crcr]{%
1	7\\
2	12\\
};
\addplot [color=black, dotted, mark=square, mark size=3pt, mark options={solid, black}, forget plot]
  table[row sep=crcr]{%
1	10\\
2	7\\
};
\end{axis}

\begin{axis}[%
width=1.5in,
height=0.5in,
at={(1.85in,0.805in)},
scale only axis,
xmin=0.5,
xmax=2.5,
xtick={1,2},
xticklabels={{3PP},{1PP}},
ymin=0,
ymax=14.5,
axis background/.style={fill=white},
title style={font=\bfseries},
title={O1},
xmajorgrids,
ymajorgrids
]
\addplot [color=black,  dotted, mark=o, mark size=3pt, mark options={solid, black}, forget plot]
  table[row sep=crcr]{%
1	5\\
2	12\\
};
\addplot [color=black, dotted, mark=square, mark size=3pt, mark options={solid, black}, forget plot]
  table[row sep=crcr]{%
1	5\\
2	6\\
};
\end{axis}

\begin{axis}[%
width=1.5in,
height=0.5in,
at={(5.845in,4.669in)},
scale only axis,
xmin=0.5,
xmax=2.5,
xtick={1,2},
xticklabels={{},{}},
ymin=0,
ymax=14.5,
axis background/.style={fill=white},
title style={font=\bfseries},
title={Fz},
xmajorgrids,
ymajorgrids
]
\addplot [color=black,  dotted, mark=o, mark size=3pt, mark options={solid, black}, forget plot]
  table[row sep=crcr]{%
1	11\\
2	8\\
};
\addplot [color=black, dotted, mark=square, mark size=3pt, mark options={solid, black}, forget plot]
  table[row sep=crcr]{%
1	11\\
2	9\\
};
\end{axis}

\begin{axis}[%
width=1.5in,
height=0.5in,
at={(5.845in,3.703in)},
scale only axis,
xmin=0.5,
xmax=2.5,
xtick={1,2},
xticklabels={{},{}},
ymin=0,
ymax=14.5,
axis background/.style={fill=white},
title style={font=\bfseries},
title={Cz},
xmajorgrids,
ymajorgrids
]
\addplot [color=black,  dotted, mark=o, mark size=3pt, mark options={solid, black}, forget plot]
  table[row sep=crcr]{%
1	14\\
2	10\\
};
\addplot [color=black, dotted, mark=square, mark size=3pt, mark options={solid, black}, forget plot]
  table[row sep=crcr]{%
1	13\\
2	14\\
};
\end{axis}

\begin{axis}[%
width=1.5in,
height=0.5in,
at={(5.845in,2.737in)},
scale only axis,
xmin=0.5,
xmax=2.5,
xtick={1,2},
xticklabels={{},{}},
ymin=0,
ymax=14.5,
axis background/.style={fill=white},
title style={font=\bfseries},
title={Pz},
xmajorgrids,
ymajorgrids
]
\addplot [color=black,  dotted, mark=o, mark size=3pt, mark options={solid, black}, forget plot]
  table[row sep=crcr]{%
1	11\\
2	11\\
};
\addplot [color=black, dotted, mark=square, mark size=3pt, mark options={solid, black}, forget plot]
  table[row sep=crcr]{%
1	13\\
2	10\\
};
\end{axis}

\begin{axis}[%
width=1.5in,
height=0.5in,
at={(5.845in,1.771in)},
scale only axis,
xmin=0.5,
xmax=2.5,
xtick={1,2},
xticklabels={{3PP},{1PP}},
ymin=0,
ymax=14.5,
axis background/.style={fill=white},
title style={font=\bfseries},
title={Oz},
xmajorgrids,
ymajorgrids
]
\addplot [color=black,  dotted, mark=o, mark size=3pt, mark options={solid, black}, forget plot]
  table[row sep=crcr]{%
1	7\\
2	0\\
};
\addplot [color=black, dotted, mark=square, mark size=3pt, mark options={solid, black}, forget plot]
  table[row sep=crcr]{%
1	6\\
2	0\\
};
\end{axis}

\begin{axis}[%
width=1.5in,
height=0.5in,
at={(9.841in,5.636in)},
scale only axis,
xmin=0.5,
xmax=2.5,
xtick={1,2},
xticklabels={{},{}},
ymin=0,
ymax=14.5,
axis background/.style={fill=white},
title style={font=\bfseries},
title={FP2},
xmajorgrids,
ymajorgrids
]
\addplot [color=black,  dotted, mark=o, mark size=3pt, mark options={solid, black}, forget plot]
  table[row sep=crcr]{%
1	7\\
2	10\\
};
\addplot [color=black, dotted, mark=square, mark size=3pt, mark options={solid, black}, forget plot]
  table[row sep=crcr]{%
1	11\\
2	10\\
};
\end{axis}

\begin{axis}[%
width=1.5in,
height=0.5in,
at={(9.841in,4.669in)},
scale only axis,
xmin=0.5,
xmax=2.5,
xtick={1,2},
xticklabels={{},{}},
ymin=0,
ymax=14.5,
axis background/.style={fill=white},
title style={font=\bfseries},
title={F4},
xmajorgrids,
ymajorgrids
]
\addplot [color=black,  dotted, mark=o, mark size=3pt, mark options={solid, black}, forget plot]
  table[row sep=crcr]{%
1	9\\
2	7\\
};
\addplot [color=black, dotted, mark=square, mark size=3pt, mark options={solid, black}, forget plot]
  table[row sep=crcr]{%
1	11\\
2	8\\
};
\end{axis}

\begin{axis}[%
width=1.5in,
height=0.5in,
at={(9.841in,3.703in)},
scale only axis,
xmin=0.5,
xmax=2.5,
xtick={1,2},
xticklabels={{},{}},
ymin=0,
ymax=14.5,
axis background/.style={fill=white},
title style={font=\bfseries},
title={C4},
xmajorgrids,
ymajorgrids
]
\addplot [color=black,  dotted, mark=o, mark size=3pt, mark options={solid, black}, forget plot]
  table[row sep=crcr]{%
1	13\\
2	9\\
};
\addplot [color=black, dotted, mark=square, mark size=3pt, mark options={solid, black}, forget plot]
  table[row sep=crcr]{%
1	12\\
2	14\\
};
\end{axis}

\begin{axis}[%
width=1.5in,
height=0.5in,
at={(9.841in,2.737in)},
scale only axis,
xmin=0.5,
xmax=2.5,
xtick={1,2},
xticklabels={{},{}},
ymin=0,
ymax=14.5,
axis background/.style={fill=white},
title style={font=\bfseries},
title={P4},
xmajorgrids,
ymajorgrids
]
\addplot [color=black,  dotted, mark=o, mark size=3pt, mark options={solid, black}, forget plot]
  table[row sep=crcr]{%
1	12\\
2	10\\
};
\addplot [color=black, dotted, mark=square, mark size=3pt, mark options={solid, black}, forget plot]
  table[row sep=crcr]{%
1	10\\
2	12\\
};
\end{axis}

\begin{axis}[%
width=1.5in,
height=0.5in,
at={(9.841in,1.771in)},
scale only axis,
xmin=0.5,
xmax=2.5,
xtick={1,2},
xticklabels={{},{}},
ymin=0,
ymax=14.5,
axis background/.style={fill=white},
title style={font=\bfseries},
title={T6},
xmajorgrids,
ymajorgrids
]
\addplot [color=black,  dotted, mark=o, mark size=3pt, mark options={solid, black}, forget plot]
  table[row sep=crcr]{%
1	9\\
2	12\\
};
\addplot [color=black, dotted, mark=square, mark size=3pt, mark options={solid, black}, forget plot]
  table[row sep=crcr]{%
1	9\\
2	13\\
};
\end{axis}

\begin{axis}[%
width=1.5in,
height=0.5in,
at={(9.841in,0.805in)},
scale only axis,
xmin=0.5,
xmax=2.5,
xtick={1,2},
xticklabels={{3PP},{1PP}},
ymin=0,
ymax=14.5,
axis background/.style={fill=white},
title style={font=\bfseries},
title={O2},
xmajorgrids,
ymajorgrids,
]
\addplot [color=black,  dotted, mark=o, mark size=3pt, mark options={solid, black}]
  table[row sep=crcr]{%
1	8\\
2	11\\
};

\addplot [color=black, dotted, mark=square, mark size=3pt, mark options={solid, black}]
  table[row sep=crcr]{%
1	5\\
2	7\\
};

\end{axis}
\end{tikzpicture}%
	}
	\caption{Comparison of the indegree (vertical axes) in $\beta$ for both conditions and participants. The indegree values for P1 and P2 are indicated with $\ocircle$ and $\Box$, respectively.}
	\label{fig:ID_beta}
\end{figure}

Similarly, Figure~\ref{fig:ID_beta} shows the comparison between 1PP and 3PP of the ID in $\beta$ for all the channels of interest of each participant. We can identify in which node a specific metric is greater or lower for both participants. For example, we can identify a greater ID in 1PP than in 3PP at FP1, which would mean that FP1 is dependent by more channels in 1PP than in 3PP. Likewise, Oz has a greater ID in 3PP than in 1PP. This means that the dependency of channel Oz from other channels is lower in 1PP than in 3PP.

\begin{figure}
	\centering
		\includegraphics[trim=0 0 0 0.43in, clip, width=0.95\textwidth]{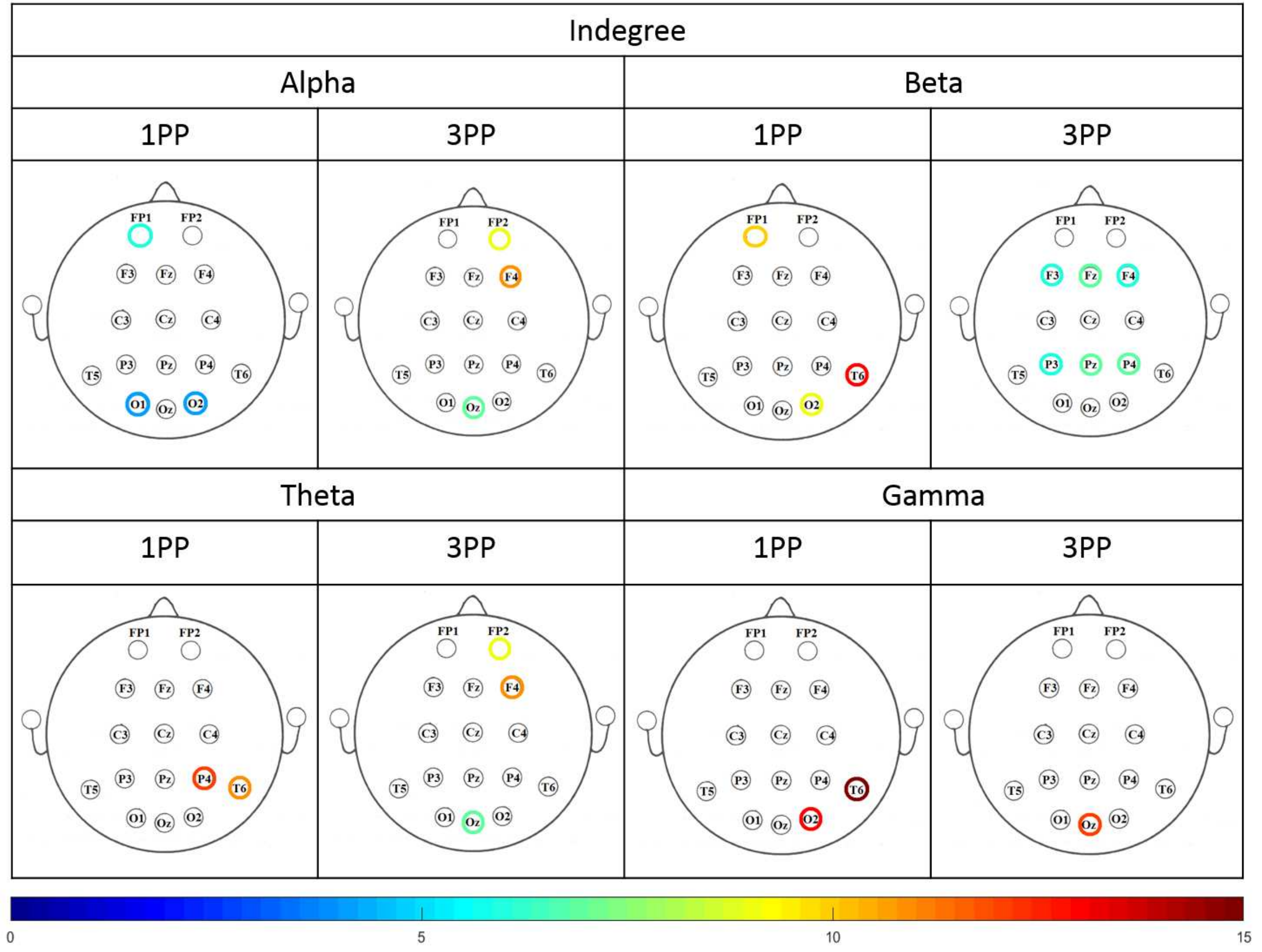}
	\caption{Mean indegree for both participants shown in color only at the sensors with an incremental trend in comparison to the other condition. For each frequency band, the head to the left shows the cases when the indegree in 1PP increased in comparison to 3PP, while the head to the right shows those for which 3PP increased in comparison to 1PP.}
	\label{fig:Indegree}
\end{figure}

\begin{figure}
	\centering
		\includegraphics[trim=0 0 0 0.43in, clip, width=0.95\textwidth]{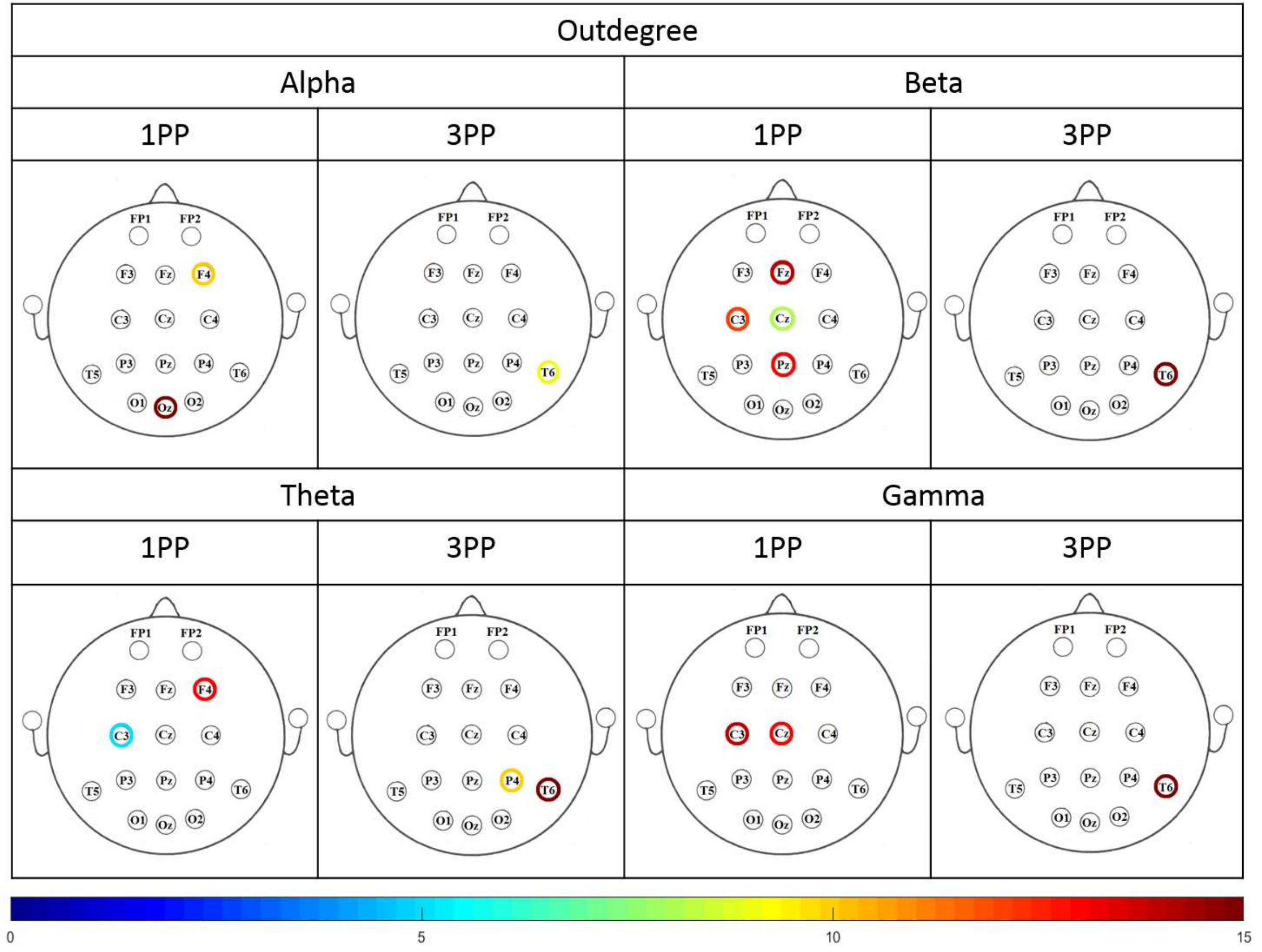}
	\caption{Mean outdegree of both participants shown in color only at the sensors with an incremental trend in comparison to the other condition. For each frequency band, the head to the left shows the cases when the outdegree in 1PP increased in comparison to 3PP, while the head to the right shows those for which 3PP increased in comparison to 1PP.}
	\label{fig:Outdegree}
\end{figure}

\begin{figure}
	\centering
		\includegraphics[trim=0 0 0 0.43in, clip, width=0.95\textwidth]{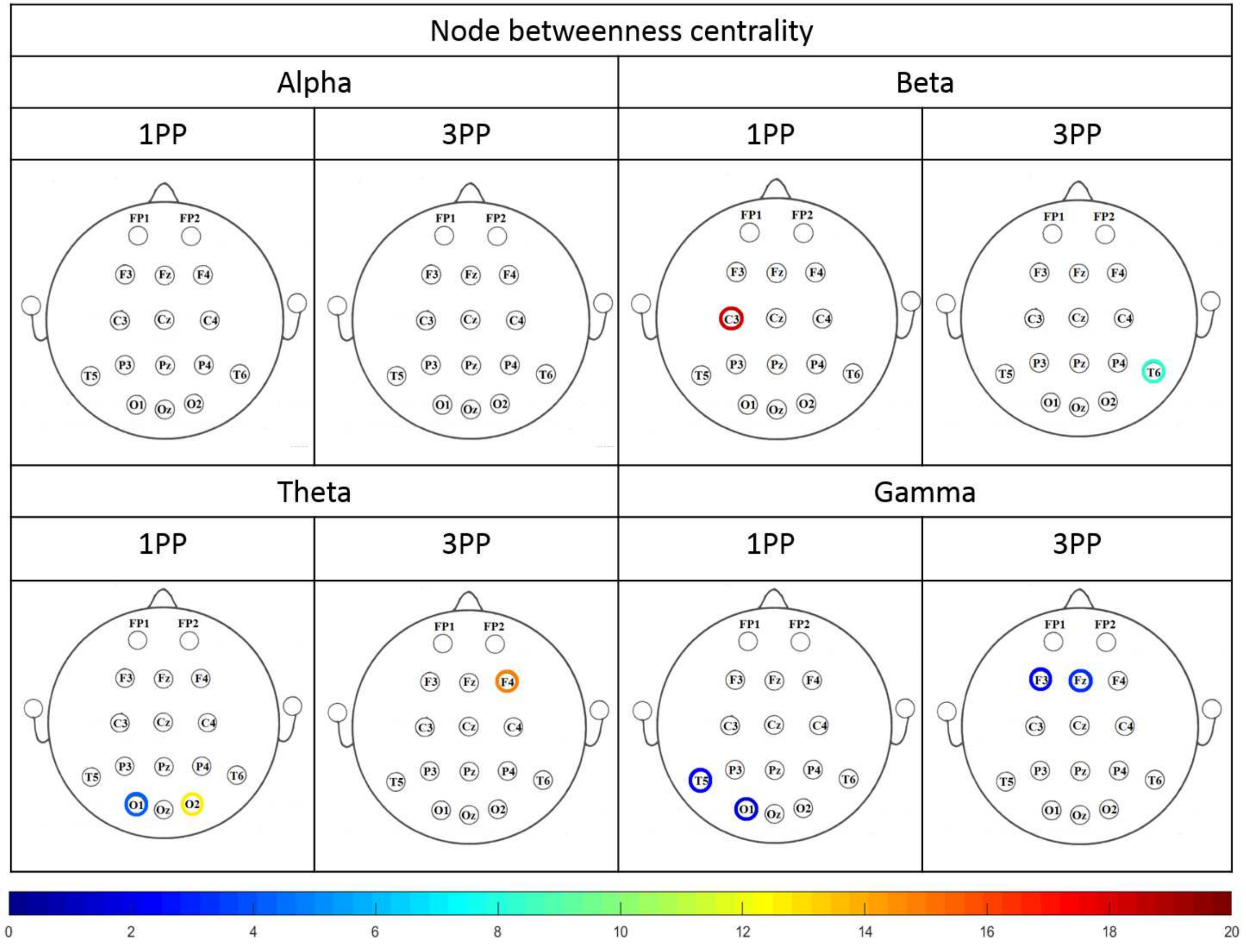}
	\caption{Mean node betweenness centrality of both participants shown in color only at the sensors with an incremental trend in comparison to the other condition. For each frequency band, the head to the left shows the cases when the node betweenness centrality in 1PP increased in comparison to 3PP, while the head to the right shows those for which 3PP increased in comparison to 1PP.}
	\label{fig:NodeBetweenness}
\end{figure}

\begin{figure}
	\centering
		\includegraphics[trim=0 0 0 0.43in, clip, width=0.95\textwidth]{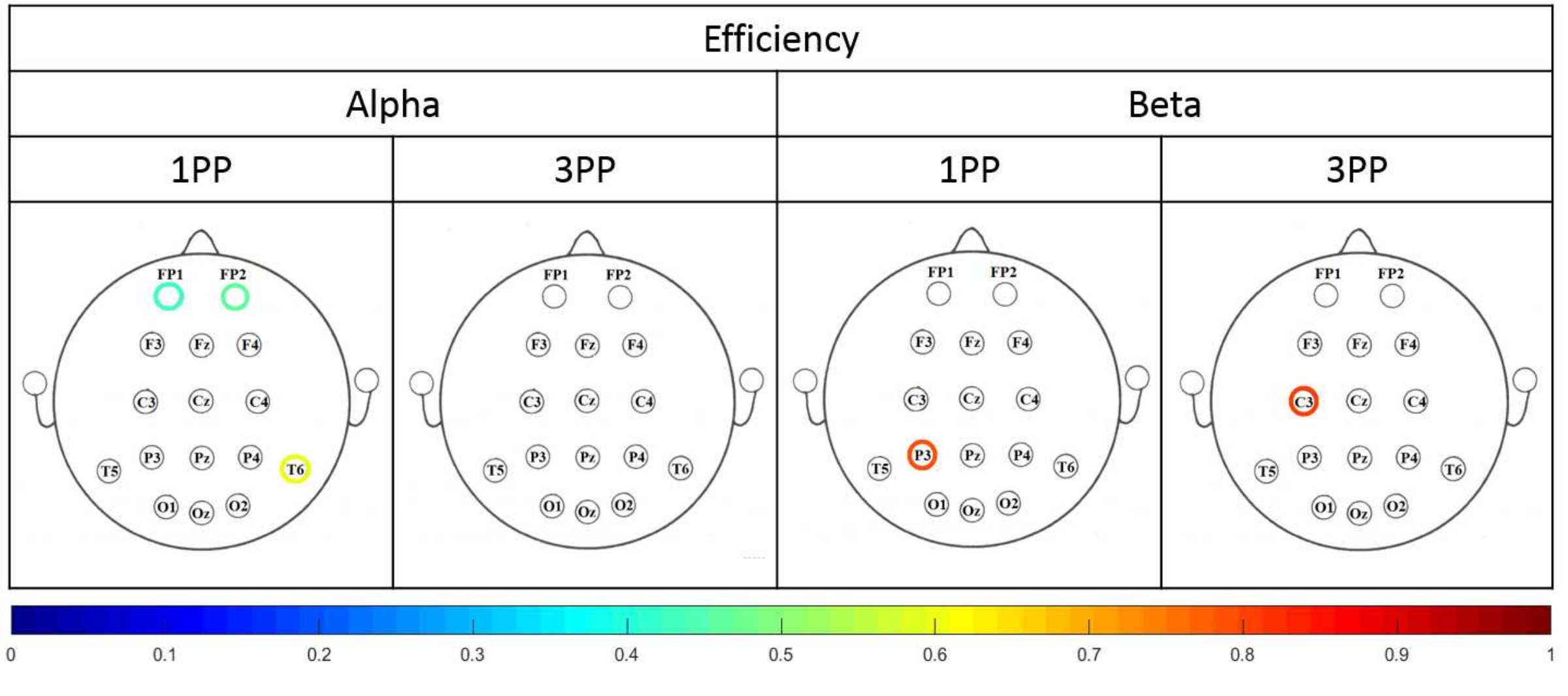}
	\caption{Mean efficiency of both participants shown in color only at the sensors with an incremental trend in comparison to the other condition. For each frequency band, the head to the left shows the cases when the efficiency in 1PP increased in comparison to 3PP, while the head to the right shows those for which 3PP increased in comparison to 1PP.}
	\label{fig:Efficiency}
\end{figure}

In Figures~\ref{fig:Indegree}, \ref{fig:Outdegree}, \ref{fig:NodeBetweenness}, and \ref{fig:Efficiency}, we show a summary of all the graph metrics obtained from our participants in the different brain rhythms. These results are presented over the array of sensors for a better view of the brain regions they correspond to. Furthermore, only the metrics that increased in both participants are shown for both 1PP and 3PP conditions:
\begin{itemize}
\item Figure~\ref{fig:Indegree}  shows the magnitude of the indegree. We can identify the stronger value of the ID in $\beta$, $\theta$ and $\gamma$ for 1PP condition at right temporal, parietal, and occipital brain regions. 
\item Figure~\ref{fig:Outdegree}, shows the magnitude of the outdegree. We can identify in $\beta$ that at C3 and the midline it is greater in 1PP than in 3PP. This could mean that the influence of these channels is stronger than the others. Additionally, there is a high OD at T6 in the 3PP condition in all frequency bands, which could mean that this channel is an important hub in 3PP condition. 
\item Figure ~\ref{fig:NodeBetweenness} shows the magnitude of the node betweenness centrality. The greatest betweenness centrality is at C3 in $\alpha$ for the 1PP condition. In $\theta$ and $\gamma$ the betweenness centrality nodes are at the temporal and occipital regions in 1PP condition. Meanwhile, for the 3PP condition are at the frontal region.  We didn't find any significant difference in $\alpha$.
\item Figure~\ref{fig:Efficiency} shows the magnitude of the local efficiency. We can identify high efficiency in $\beta$ for both conditions, but for 1PP is greater at P3, while from 3PP is greater at C3. We did not find any significant differences in the other frequency bands. As an important result, we identified a greater efficiency in 1PP than in 3PP at the prefrontal area (FP1 and FP2). This could be due to the sense of agency, similarly to \cite{Baumgartner.2008} where they found the prefrontal areas strongly involved in the modulation of experience of presence in adults. 
\end{itemize}
\section{Discussion}
In this proof-of-concept study, the use of graph theory allowed us to investigate relevant features in brain networks through EEG data. Such data was acquired from an experimental BCI system that is well suited to describe the differences of sense of agency in two environments with different levels of immersion.
We calculated functional connectivity based on PDC because it is a metric that allows for an analysis in the frequency domain, then giving us more information about the cognitive processes in a specific band. This brain network analysis could help us understand the causal relationships and give us an idea of the information flow and brain organization during the control of an immersive BCI and sense of agency. 

Our results show the frontal and parietal brain regions as target areas during 3PP, mainly in $\beta$. This is consistent with~\cite{Rathee.2016} who reported that, using the partial Granger causality during the right hand MI, the higher amount of outgoing information was from the central region (C3, Cz, C4) and the targets were Fz and Pz. %
Furthermore, the nodes with higher ID in 3PP in $\alpha$ and $\beta$ are in accordance with the results in~\cite{Athanasiou.2018}, where the nodes with higher in-degree are associated with the supplementary motor areas bilaterally during hand motor imagery, which corresponds to F3, Fz, and F4. Moreover, $\alpha$ and $\beta$ presents a high ID at FP1 in 1PP condition, which might possibly be related to a better coordinative role of these rhythms in the planning of the sensorimotor process. Also, \cite{Baumgartner.2008} found that the prefrontal areas are strongly involved in modulating the experience of presence. %
In~\cite{Ghosh.2015}, they found as characteristic the out strength of C3 while performing motor imagery of the right hand. In our case, we found that the OD at C3 in 1PP is higher than in 3PP. Then, the stronger influence from C3 to other regions might be possibly a reflection of a better engagement. 

In this study we found a greater OD at the SMA (C3, Cz) in 1PP than in 3PP in $\beta$ and $\gamma$, which could be related to the regulative role of the SMAs during the planning of a movement, as identified in~\cite{Athanasiou.2012}. Furthermore, when performing MI of right hand in 1PP, there is a greater OD and node betweenness centrality  in $\beta$ than in 3PP, which can be an indication of the existence of a central node for information exchange. This is in line with results from~\cite{Li.2019b}.

The node betweenness centrality shows us an important functional connection at C3 in $\beta$, which could mean that C3 plays the role of regulator in the flow of information during the MI in 1PP condition. 
As seen in $\alpha$, the efficiency is greater in 1PP condition than in 3PP at the prefrontal cortex, which could be considered as a structural candidate 
for modulating inter-individual differences in the experience of presence~\citep{Baumgartner.2008}.

The present study is limited by its small sample size in human subjects, but our primary goal was the development and demonstration of our experimental BCI system and the analysis approach of the data we collected from it. Future work would include a more exhaustive experimentation in different subjects, as well as increasing the sense of agency by incorporating other activities and other interfaces, such as the one described by~\cite{berlin19}. Additionally, and given that in this work we only analyzed the data during the MI epochs, we would like to extend the analysis to the feedback epochs as well so to discard the possible involvement of mirror system neurons~\citep{pineda2005functional} and properly assess the experience of sense of agency. 

\section{Concluding Remarks}
In this paper, we showed the applicability of an experimental BCI system for the interaction with a robot through different immersion experiences. The BCI system we proposed can be seen as a tool for the analysis of functional brain connectivity associated to the control of a BCI in an immersive environment. For that purpose, we expended our previous work on connectivity measures based on the PDC by using metrics of graph theory. As a proof-of-concept, we analyzed the data from two participants, and the results seem to be in line with previous results from the literature.
 

\section*{Conflict of Interest Statement}
The authors declare that the research was conducted in the absence of any commercial or financial relationships that could be construed as a potential conflict of interest.

\section*{Author Contributions}
AM and GD conceived and designed the study. AM collected the data. AM performed the signal processing and statistical analysis. AM contributed to the interpretation of the findings. AM drafted the paper. AM and GD proofread and edited the paper. All authors read and approved the final version of the paper.

\section*{Funding}
Declared none.

\section*{Acknowledgments}
Declared none.

\end{document}